\documentclass[aps,amssymb,amsmath,amsfonts,twocolumn,pra]{revtex4}
\usepackage[latin1]{inputenc}
\usepackage{amsthm} 
\usepackage{amsmath}
\usepackage{graphicx}
\usepackage{subfigure}
\usepackage{amsfonts}
\usepackage{hyperref}
\usepackage{xcolor}
\newtheorem{cons}{CONSTRUCTION}

\begin{document}
\title{Multipartite entanglement in heterogeneous systems}
\author{Dardo Goyeneche}
\affiliation{Faculty of Physics, Warsaw University, Pasteura 5, 02-093 Warsaw, Poland}
\affiliation{Faculty of Applied Physics and Mathematics, Technical University of Gda\'{n}sk, 80-233 Gda\'{n}sk, Poland}
\affiliation{Institute of Physics, Jagiellonian University, Krak\'ow, Poland}
\author{Jakub Bielawski}
\affiliation{Faculty of Physics, Astronomy and Computer Sciences, Jagiellonian University, Krak\'ow, Poland}
\affiliation{Cracow University of Economics, Krak\'ow, Poland}
\author{Karol {\.Z}yczkowski}
\affiliation{Institute of Physics, Jagiellonian University, Krak\'ow, Poland}
\affiliation{Center for Theoretical Physics, Polish Academy of Sciences, Warsaw, Poland}

\date{May 12th, 2016}

\begin{abstract}
Heterogeneous bipartite quantum pure states, composed of two subsystems with a different number of levels, cannot have both reductions maximally mixed. In this work, we demonstrate existence of a wide range of highly entangled states of heterogeneous multipartite systems consisting of $N>2$ parties such that every reduction to one and two parties is maximally mixed. Two constructions of generating genuinely multipartite maximally entangled 
states of heterogeneous systems for an arbitrary number of subsystems are presented. Such states are related to quantum error correction codes over mixed alphabets and mixed orthogonal arrays. Additionally, we show the advantages of considering heterogeneous systems in practical implementations of multipartite steering.
\end{abstract}
\maketitle

\section{Introduction}
Characterization of entanglement in multipartite quantum systems is an important open problem in quantum theory. Entangled states have a fundamental role in quantum teleportation, quantum key distribution, dense coding and error correcting codes \cite{S04,Arnaud} and quantum computation \cite{Jozsa}. Furthermore, they allow us to generate multipartite Bell inequalities \cite{Guhne} and to formulate an independent proof of the Kochen-Specker theorem \cite{Cabello}. In recent years, multipartite entanglement for homogeneous systems of $N$-qudit systems, i.e., pure states belonging to a Hilbert space $\mathcal{H}_d^{\otimes N}$, has been thoroughly studied \cite{S04}. In particular, graph states \cite{Hein} provide many classes of entangled states, including the maximally entangled ones.

In a more general setup, quantum states belong to a Hilbert space of the form 
$\mathcal{H}_{d_1}^{\otimes n_1}\otimes\mathcal{H}_{d_2}^{\otimes n_2}\otimes\dots\mathcal{H}_{d_l}^{\otimes n_l}$,
 where the local Hilbert spaces $\mathcal{H}_{d_1},\mathcal{H}_{d_2},\dots,\mathcal{H}_{d_l}$ have 
possibly different dimensions. Such a Hilbert space contains
 quantum states associated to \emph{heterogeneous} systems, 
having different number of levels. For brevity we
 will refer to such systems as $d_1^{n_1}\times d_2^{n_2}\times\dots\times d_l^{n_l}$. 
Investigation of entanglement in heterogeneous systems was recently performed
in several particular cases, e.g. for tri-partite systems, 
$2 \times 2 \times n=2^2\times n$ \cite{YSW07,YZS08} and
$2\times n_1\times n_2$  \cite{Miyake,CC06,Chen2},
and for four-partite systems, $2^3\times n$ \cite{WLL13}. 
A key problem for characterizing entanglement in heterogeneous systems is the lack of a suitable mathematical tool -- for instance the Galois fields do not exist in non-prime power dimensions. Therefore, the study of entanglement for heterogeneous systems is even more challenging than for the homogeneous case.

Properties of quantum entanglement for heterogeneous bipartite systems are well understood, whereas the complexity of the analogous problem for tripartite systems becomes intractable in general.
It has been proven that the problem of calculating the rank of a tensor with three indices 
over any finite field is NP-complete with respect to the dimension of the tensor \cite{Hastad}.

The codification of information in the orbital angular momentum of photons is a resource to generate an unbounded number of discrete levels \cite{TT11,AB12}. This technique has allowed one to improve classical \cite{W12,B13} and quantum \cite{GJVWZ06,MMORMGB15} communication protocols. Remarkably, 
 taking into account subsystems consisting of more than two levels each, improves security of some quantum information protocols \cite{L04,M04,G06,C02}, 
increases capacity of quantum channels \cite{F03} and efficiency of quantum gates \cite{R07}. Recently, experimentalists paid attention to the entanglement of different degrees of freedom (e.g. path-polarization) and successfully generated a \emph{qubit-qutrit} hyperentangled state \cite{LWLBRGW08}. Additionally, a three partite state composed of one \emph{qubit}  and two  \emph{qutrits} exhibiting  genuinely multipartite entanglement was generated in laboratory \cite{MEHKFZ15}. These results provide a motivation to study protection of entanglement in heterogeneous systems under decoherence \cite{XL13,X14}.  It is fair to expect that multipartite entangled states of heterogeneous systems consisting of several qubits and qutrits will be implemented in a near future.  However, the theory of quantum entanglement in heterogeneous systems consisting of several subsystems of different sizes is by far not complete.
The main aim of this work is to provide a concrete contribution in this direction by considering mixed orthogonal arrays.

A pure mutipartite state of a system containing $N$ subsystems
is called $k$--uniform if every density matrix reduced to $k$ subsystems is maximally mixed \cite{S04}.
Such states are also called maximally multipartite entangled  \cite{Facchi}
and can be considered as multipartite generalizations of the 
two--qubit Bell state. 
Maximally entangled states of homogeneous systems consisting of $N$
subsystems of $d$ levels each are closely related to
quantum error correcting codes
for messages encoded in an alphabet consisting of $d$ letters \cite{S04,Arnaud}.
Furthermore, such states are useful 
for quantum secret sharing protocol \cite{HCLRL12} 
and to design holographic quantum codes \cite{LS15,PYHP15,ZHQ15}.

Highly entangled $k$--uniform multipartite states can be constructed 
from graph states \cite{He13},
orthogonal arrays \cite{GZ14},
mutually orthogonal Latin squares and Latin cubes \cite{GALRZ15}
and symmetric matrices \cite{FJXY15}.
In the present paper, we consider an extension of orthogonal arrays, known as 
\emph{mixed orthogonal arrays} (MOA) \cite{HSS99},
which allows us to construct genuinely multipartite maximally entangled states
 for heterogeneous systems,
 related to  quantum error correction codes over mixed alphabets \cite{WYFO12}.

 This work is organized as follows: In Section II we establish a link between mixed 
orthogonal arrays and multipartite entanglement for heterogeneous systems. 
Additionally, the central concept of \emph{irredundant MOA} is introduced together 
with basic properties and clarifying examples.
 In Section III, two powerful constructions to generate irredundant MOA are shown. These constructions allow us to generate maximally entangled states for a wide range of heterogeneous systems. In particular, we prove the existence of 1-uniform and 2-uniform states for heterogeneous systems for any number of parties. In Section IV, we exemplify our constructions by presenting explicitly some new quantum states. Furthermore, we show that every quantum state generated by our proposed techniques can be written as linear combination involving orthogonal $GHZ$-like states. In Section V we introduce the concept of \emph{endurance} of $k$ uniformity, which allows us to extend the variety of $k$-uniform states that we find. In Section VI, we present some applications of maximally entangled states for heterogeneous systems to quantum steering. Furthermore, we construct the simplest 1-uniform state of this work corresponding to a system of two qubits and one qutrit. In Section VII, we discuss main results of the work and present some open questions. Finally, in Appendix \ref{AppendixA} we show how to find maximally entangled states from redundant MOA and we exemplify this method by finding a 1-uniform state consisting of one qubit and two qutrits, which is not possible to find from an irredundant MOA.

\section{Heterogeneous entanglement and mixed orthogonal arrays}\label{S2}
Orthogonal arrays provide  a fundamental tool to design experiments and 
are known as \emph{Taguchi Designs} \cite{Roy}. In a previous work we studied orthogonal arrays and their relation to maximally entangled states of homogeneous systems consisting of $N$ subsystems with $d$ levels each \cite{GZ14}. Here, we consider \emph{ mixed orthogonal arrays} \cite{HSS99,Rao}, 
also called asymmetric orthogonal arrays \cite{HPS92}, 
which form a natural generalization of orthogonal arrays (OA). A mixed orthogonal array MOA$(r,d_1^{n_1}, d_2^{n_2}\dots d_l^{n_l},k)$ is an array of $r$ rows and $N$ columns ($r\times N$ array), 
with $N=n_1+n_2+\dots+n_l$ such that the first $n_1$ columns have symbols from $\{0,1,\dots d_1-1\}$, the next $n_2$ columns have symbols from $\{0,1,\dots d_2-1\}$ and so on, with the property that any $r\times k$ subarray contains every possible combination of $k$ symbols with the same number of appearance. 
An OA is a particular case of MOA having identical set of symbols in every column, that is, $d_1=d_2=\dots=d_l$. The notation considered here for OA and MOA was introduced by Rao \cite{Rao}.
For an excellent introduction to orthogonal arrays and their applications
consult the book of Hedayat, Sloane and Stufken \cite{HSS99}. 

Orthogonal arrays offer a convenient tool to generate highly 
entangled states for homogeneous multipartite systems composed of $N$ qudits \cite{GZ14},
including Dicke states \cite{D54}, cluster states \cite{BR01} and  graph states \cite{HEB04}. A MOA$(r,d_1^{n_1},\dots,d_l^{n_l},k)$ having $N=n_1+\dots+n_l$ columns is 
called \emph{irredundant}, written IrMOA,
if every subset of $N-k$ columns contains a different sequence of $N-k$ symbols in every row. It was demonstrated in \cite{GZ14} that an IrOA($r,d^N,k$) leads to a $k$-uniform state of $N$ subsystems with $d$ levels each. In the present work, we extend the same idea to the mixed alphabets by considering IrMOA($r,d_1^{n_1},d_2^{n_2},\dots,d_l^{n_l},k$)
\begin{equation}\label{array}
\begin{array}{cccc}
    a_{1,1} & a_{1,2} & \dots & a_{1,N} \\
    a_{2,1} & a_{2,2} & \dots & a_{2,N} \\
    \vdots & \vdots & \dots& \vdots \\
    a_{r,1} & a_{r,2} & \dots & a_{r,N}
  \end{array},
\end{equation}
which lead to $k$--uniform states of the system $d_1^{n_1}\times d_2^{n_2}\times\dots\times d_l^{n_l}$, consisting of $n_1$ subsystems having $d_1$ levels, $n_2$ subsystems having $n_2$ levels and so on. Without loss of generality we assume that $d_1>d_2>\dots>d_l$. The generated state, denoted as 
$|\phi_{d_1^{n_1},d_2^{n_2},\dots,d_l^{n_l}}\rangle$,
 is of the form
\begin{eqnarray}\label{pure2}
&|a_{1,1},a_{1,2},\dots,a_{1,N}\rangle+&\nonumber\\
&|a_{2,1},a_{2,2},\dots,a_{2,N}\rangle+&\nonumber\\
&\vdots&\nonumber\\
&|a_{r,1},a_{r,2},\dots,a_{r,N}\rangle{\color{white}+}.&
\end{eqnarray}
For brevity the normalization factors in front of pure states discussed in this work are omitted.
In particular, an IrOA($r,d^N,k$), denoted IrOA($r,N,d,k$) in \cite{GZ14}, defines a $k$-uniform state of a homogeneous system consisting of $N$ subsystems with $d$ levels each.
This condition means that the partial trace over any selected $N-k$ subsystems
forms the maximally mixed state of $k$ qudits.

The classes of states that can be constructed from IrMOA posses
two remarkable properties \cite{GZ14}: \emph{(A) Uniformity:} every combination of $k$ symbols has the same number of appearance along the rows. This implies that every reduction to $k$ qudits, $\rho_k=\mathrm{Tr}_{N-k}(|\Phi\rangle\langle\Phi |)$, has a uniform diagonal. \emph{(B) Diagonality:} the irredundancy of the MOA implies that every reduction  $\rho_k$ to $k$ qudits forms a diagonal matrix. In this way, OA and irredundancy are the two conditions sufficient to assure uniformity and diagonality of partial traces. This in turn implies that the corresponding pure state is $k$--uniform. We remark that the same properties hold in the same way for states constructed from IrMOA. Let us present some examples. The array
\begin{equation}\label{array2}
\mbox{IrOA}(2,2^2,1)=
\begin{array}{cc}
0&0\\
1&1
 \end{array},
 \end{equation}
defines the standard Bell state, $|00\rangle+|11\rangle$, while 
 \begin{equation}
\mbox{IrOA}(2,2^3,1)=
\begin{array}{ccc}
0&0&0\\
1&1&1
  \end{array},
\end{equation}
yields the GHZ state, $|000\rangle+|111\rangle$.  

In the case of multilevel systems we have, for instance
\begin{equation}\label{MOA2231}
{\rm IrMOA}(4,4^12^2,1)=\begin{array}{ccc}
0&0&0\\
1&0&1\\
2&1&0\\
3&1&1
\end{array},
\end{equation}
which defines a $1$--uniform state of
a system $4\times2^2$ consisting of one ququart and two qubits,
\begin{equation}
\label{AME2231}
|\phi_{4^12^4}\rangle=|000\rangle+|101\rangle+|210\rangle+|311\rangle.
\end{equation} 
The MOA (\ref{MOA2231}) is listed in the on-line catalog of MOA of strength 2 provided by Kuhfeld \cite{Kuhfeld}. However, state (\ref{AME2231}) does not capture the aim of our goal. Note that the ququart in the state (\ref{AME2231}) can be decomposed into two qubits in order to get a 1-uniform state of 4 qubits systems:
\begin{equation}
|\phi_{2^4}\rangle=|0000\rangle+|0101\rangle+|1010\rangle+|1111\rangle,
\end{equation} 
where we considered the identification: $|0\rangle\rightarrow |00\rangle$, $|1\rangle\rightarrow |01\rangle$, $|2\rangle\rightarrow |10\rangle$ and $|3\rangle\rightarrow |11\rangle$.
 By this reason, in the rest of the work we will be mainly focused on \emph{genuinely heterogeneous systems},
 i.e., systems composed of subsystems with coprime levels (e.g. qubits-qutrits or qutrits-ququints). 
These kind of heterogeneous systems cannot be transformed into homogeneous 
systems by an identification of symbols, like the one used
for the state (\ref{AME2231}).

\section{Construction of Irredundant OA and MOA}\label{S_IrMOA}
In this section, we provide a general framework to construct IrMOA which represents a natural 
tool to generate genuinely multipartite entangled states for heterogeneous systems 
with an arbitrary number of parties. In the previous work \cite{GZ14}
we used the fact that real Hadamard matrices allow one to construct IrOA($r,2^N,2$) for $N>5$ and 
suitable values of $r$. From such arrangements we were able to generate 2-uniform states for $N>5$ qubits systems. Here, we generalize this result to IrMOA by considering the combinatorial notion of \emph{difference schemes} \cite{HSS99,B39,CD96}. A difference scheme $D(s,N,d)$ is an arrangement having $s$ rows, $N$ columns and $d$ different symbols such that the difference between every pair of rows, with respect to operations in the \emph{Galois field} $GF(d)$, contains the $d$ symbols equally often. Thus, there arises the following method to obtain IrMOA of strength one from difference schemes.
\begin{cons}\label{C1}
If a difference scheme $D(s,N,d)$ exists then the IrMOA($ds,d^Np^1_1\dots p^1_m,1$) exists, where $p_1,\dots,p_m$ are the $m$ distinct prime factors of $ds$.
\end{cons}
The construction of the IrMOA consists in two steps. First, we consider the
difference scheme $D(s,N,d)$ and the following juxtaposition:
\begin{eqnarray}
&D(s,N,d)+0&\nonumber\\
&D(s,N,d)+1&\nonumber\\
&\vdots&\nonumber\\
&D(s,N,d)+d-1,&
\end{eqnarray}
where $D(s,N,d)+j$ means that every entry of the difference scheme $D(s,N,d)$ is increased by $j$. Note that the resulting array have to contain $d$ different symbols in $GF(d)$. For example, if $d$ is prime then the symbols of the arrangement are numbers in $\mathbb{Z}_d$ and the sum is taken modulo $d$. See Appendix \ref{AppendixB} for the construction when $d$ is prime power. This process generates an IrOA($ds,d^N,1$). Second, in order to obtain the IrMOA($ds,d^Np^1_1\dots p^1_m,1$) we have to add $m$ columns $C^1,\dots,C^m$ with entries $(C^l)_j=j\mod p_l$, where $j=1,\dots,ds$ and $l=1,\dots,m$. 

Let us construct an explicit example from $D(2,2,2)=\{00;01\}$ with help of Construction \ref{C1}. We obtain then the IrMOA($4,2^22^1,1$) because the only prime factor of 4 is 2. This IrMOA is in fact the IrOA($4,2^3,1$), namely
\begin{equation}
\label{IrOA222}
{\rm IrOA}(4,2^3,1)=
\begin{array}{cccc}
0&0&0\\
0&1&1\\
1&1&0\\
1&0&1
\end{array}.
\end{equation}
In order to construct a \emph{genuine IrMOA} we have to consider a difference scheme $D(s,N,d)$ such that $ds$ is not the integer power of a prime number. For example, we can consider $D(2,3,3)=\{000;012\}$ 
and obtain the genuinely mixed orthogonal array IrMOA($6,3^33^12^1,1$)$\equiv$IrMOA($6,3^42^1,1$),
\begin{equation}
\label{IrMOA3321}
 {\rm IrMOA} (6,3^42^1,1)=
\begin{array}{cccccc}
0&0&0&0&0\\
0&1&2&1&1\\
1&1&1&2&0\\
1&2&0&0&1\\
2&2&2&1&0\\
2&0&1&2&1
\end{array}.
\end{equation}
Let us present a further example. We shall skip the case $D(2,4,4)=\{0000;0123\}$ as $ds=4\times2=8$ defines an IrOA instead of a genuine IrMOA. So, we discuss the larger case, $D(2,5,5)=\{00000;01234\}$, which leads to mixed orthogonal array IrMOA($10,5^55^12^1,1$)$\equiv$IrMOA($10,5^62^1,1$). That is
\begin{equation}
\label{IrMOA5521}
{\rm IrMOA}(10,5^62^1,1) =
\begin{array}{ccccccc}
0&0&0&0&0&0&0\\
0&1&2&3&4&1&1\\
1&1&1&1&1&2&0\\
1&2&3&4&0&3&1\\
2&2&2&2&2&4&0\\
2&3&4&0&1&0&1\\
3&3&3&3&3&1&0\\
3&4&0&1&2&2&1\\
4&4&4&4&4&3&0\\
4&0&1&2&3&4&1 
\end{array}.
\end{equation}
The above examples can be generalized to any number of columns $N$ by considering the difference scheme 
\begin{equation}\label{DiffSche}
D(2,N,N)=\{0,0,\dots,0;0,1,\dots,N-1\}.
\end{equation}

\medskip

Until now we constructed irredundant mixed orthogonal arrays of strength $k=1$ from difference schemes of the form D($2,N,N$). However, Construction 1 also works for more general kinds of difference schemes. Here, we have a remarkable observation: From considering Construction \ref{C1} and a difference scheme of the form  D($N,N,d$) we produce a MOA of strength 2 \cite{B39} that is irredundant. This property has a central role in this study and it represents the most important contribution of the present paper. We consider this special result as a separate construction:
\begin{cons}\label{C2}
If a difference scheme $D(N,N,d)$ exists for $N>2$ then an ${\rm IrMOA}(dN,d^Np^1_1\dots p^1_m,2)$ exists, where $p_1,\dots,p_m$ are the $m$ distinct prime factors of $dN$ and $d$ is a prime power number.
\end{cons}
The case $N=2$ implies $d$=2 and it does not work here because the number of columns of the resulting IrMOA is three and this array is too small to assure irredundancy. This case generates the 1-uniform state of Eq.(\ref{IrOA222}). It is important to note that if $d$ is prime then the IrMOA arising from Construction \ref{C2} are given in the same way as Construction \ref{C1}. However, if $d$ is a prime power then we have to consider the Galois field $GF(d)$ (see Appendix \ref{AppendixB}). Otherwise the IrMOA only has strength $k=1$.

It worth noting that a class of difference schemes of the form $D(p,p,p)$ can be generated from the Fourier matrix
 $(F_p)_{jk}=\omega_p^{jk}=\omega_p^{D(p,p,p)_{jk}}$ of prime dimension $d=p$, where $\omega_p=e^{2\pi i/p}$ and $D(p,p,p)_{jk}$ denotes the entry of $D(p,p,p)$ located at the $j$-th row and $k$-th column. Furthermore, $D(p^m,p^m,p)$ is generated from the tensor product of $m$ Fourier matrices $(F^{\otimes m}_p)_{jk}=\omega_p^{D(p^m,p^m,p)_{jk}}$. Let us present some examples of Construction \ref{C2} by considering some of these difference schemes. From $D(3,3,3)=\{000;012;021\}$ we note that there is a single prime factor of $dN=9$, so we can generate the IrMOA($9,3^33^1,2$)$\equiv$IrOA($9,3^4,2$). That is,
\begin{equation}
\label{IrOA9432}
{\rm IrOA}^t(9,3^4,2) =
\begin{array}{ccccccccc}
0&0&0&1&1&1&2&2&2\\
0&1&2&1&2&0&2&0&1\\
0&2&1&1&0&2&2&1&0\\
0&1&2&0&1&2&0&1&2
\end{array},
\end{equation}
where $t$ denotes transposition. 
Analogously, from difference scheme
$D(4,4,4)=\{0000;0123;0231;0312\}$ we get IrMOA($16,4^42^1,2$) given by
\begin{equation}
\label{IrOA16442}
\begin{array}{cccccccccccccccc}
0&0&0&0&1&1&1&1&2&2&2&2&3&3&3&3\\
0&1&2&3&1&0&3&2&2&3&0&1&3&2&1&0\\
0&2&3&1&1&3&2&0&2&0&1&3&3&1&0&2\\
0&3&1&2&1&2&0&3&2&1&3&0&3&0&2&1\\
0&1&0&1&0&1&0&1&0&1&0&1&0&1&0&1
\end{array}.
\end{equation}
See Appendix \ref{AppendixB} for an explicit construction. Here, this IrMOA can be transformed into IrOA($16,2^9,2$). Therefore, examples (\ref{IrOA9432}) and (\ref{IrOA16442}) lead us to OA instead of genuinely mixed OA. The simplest genuine IrMOA arises for difference scheme $D(6,6,3)$, as $dN=18$ is not a power of a prime. This difference scheme can be found in the online catalog of Kuhfeld \cite{catalog_Kuhfeld}. To save the space we will not write here the resulting IrMOA($18,3^63^12^1$)$\equiv$IrMOA($18,3^72^1$), but in Section \ref{S_QS} we construct the associated quantum state.

We would like to stress that orthogonal arrays are simple to construct from difference schemes $D(s,N,d)$ when the number of symbols $d$ is prime. In this case, the symbols to be considered are elements of the group $\mathbb{Z}_d$ and the difference operation is taken modulo $d$, as we showed along the work. However, when $d$ is a prime power the construction of orthogonal arrays from difference schemes requires to use Galois fields. In Appendix \ref{AppendixB} we exemplify the construct with Galois fields by constructing the irredundant orthogonal array OA(16,4,4,2), and the corresponding quantum state, from the difference scheme $D(4,4,4)$. 

\section{Quantum states from irredundant MOA}
\label{S_QS}
Before constructing $k$-uniform states related to IrMOA it is interesting to study the maximal values of the strength $k$ allowed for systems $d_1^{n_1}\times\dots\times d_l^{n_l}$. For homogeneous systems consisting of $N$ qudits having $d$ levels each the upper bound is $k\leq N/2$. The states saturating this inequality are called {\sl absolutely maximally entangled} (AME) \cite{HCLRL12,HC13}. For example, Bell states are AME for two qubit systems. For a general $d_1 \times d_2$ system with $d_1\ne d_2$ AME states do not exist. This is because the von Neumann entropies of the reduced density matrices
 $\rho_1={\rm Tr}_2(\rho_{12})$ and $\rho_2={\rm Tr}_1(\rho_{12})$ are equal, so that both reductions of different dimensions cannot be maximally mixed. Following the same argument, $k$-uniform states consisting of $N=2k$ heterogeneous subsystems do not exist \cite{S04}. Despite this fact, there is a place for a wide range of $k$-uniform states with $k<N/2$. To get a more precise upper bound for $k$ note that for a system $d_1\times\dots\times d_{l}$, 1-uniform states do not exist if $d_1$ is larger than the dimension of the complementary system, i.e., $d_2d_3\dots d_l$. In general, a $k$-uniform state do not exist if the product of the size of $k$ local Hilbert spaces is larger than the dimension of the complementary system. This result can be stated as follows: a necessary condition for the existence of a $k$ uniform state of a system $d_1^{n_1}\times\dots\times d_l^{n_l}$  is
\begin{equation}\label{ineq}
\left(\prod_{i=1}^kd_i^{n^{\prime}_i}\right)^{2}\!\leq\,\,\prod_{i=1}^ld_i^{n_i},
\end{equation}
where $n^{\prime}_i=\min\{n_i,\max\{k-\sum_{j=1}^{i-1}n_j,0\}\}$. Here, we recall the already assumed convention, $d_1>d_2>\dots>d_l$. In the particular case of $N$ qudits with $d$ levels each Eq.(\ref{ineq}) reduces to the standard bound, $k\leq N/2$. Furthermore, a $k$-uniform state consisting of $n_1\geq k$ qutrits and $n_2$ qubits satisfies the bound
\begin{equation}
 3^{2k}\leq3^{n_1}2^{n_2},
 \end{equation}
for any $n_2$. Let us now to construct 1-uniform and 2-uniform states applying Constructions \ref{C1} and \ref{C2}, respectively. We start by considering the 1-uniform state of 3 qubits 
arising from IrOA($4,2^3,1$) 
of Eq.(\ref{IrOA222}) which is accidentally homogeneous,
\begin{equation}
|\phi_{2^3}\rangle=|000\rangle+|011\rangle+|110\rangle+|101\rangle.
\end{equation}
Here, we followed the identification between IrMOA and quantum states given in Eq.(\ref{array}) and Eq.(\ref{pure2}), where states are not normalized. A genuinely heterogeneous maximally entangled state arises from the array
IrMOA($6,3^4 2^1,1$) of Eq.(\ref{IrMOA3321}),
 which allows us to obtain the 1-uniform state of the system $3^4\times2$:
\begin{eqnarray}\label{phi3421}
|\phi_{3^42^1}\rangle&=&|00000\rangle+|01211\rangle+|11120\rangle+\nonumber\\
&&|12001\rangle+|22210\rangle+|20121\rangle.
\end{eqnarray}
Additionally, IrMOA($10,5^62^1,1$), 
explicitly written in (\ref{IrMOA5521}) 
yields the 1-uniform state of the system $5^6\times 2$,
\begin{eqnarray}\label{phi5521}
|\phi_{5^6 2^1}\rangle&=&|0000000\rangle+|0123411\rangle+|1111120\rangle+\nonumber\\
&&|1234031\rangle+|2222240\rangle+|2340101\rangle+\nonumber\\
&&|3333310\rangle+|3401221\rangle+|4444430\rangle+\nonumber\\
&&|4012341\rangle.
\end{eqnarray}
States (\ref{phi3421}) and (\ref{phi5521}) can be generalized 
for any number of parties by considering the difference scheme $D(2,N,N)$ defined in Eq.(\ref{DiffSche}). Thus, we generate the following 1-uniform state,
\begin{eqnarray}
|\phi_{N^N N}\rangle&=&\sum_{j=0}^{1}\sum_{l=0}^{d-1}\bigotimes_{k=0}^{N-1}|D(2,N,N)_{jk}+l\rangle\otimes\nonumber\\
&&\hspace{1cm}|[2l+j]_{p_1}\dots [2l+j]_{p_m}\rangle,
\end{eqnarray}
where $[X]_a=X\,(\mathrm{mod}\,a)$ and $p_1,\dots, p_m$ are the $m$ distinct prime factors of $2N$. Furthermore, 
Construction \ref{C2} allows us to generate 2-uniform states. 
The simplest cases arise from IrOA($9,3^4,2$), IrMOA($16,4^42^1,2$)$\equiv$IrOA($16,2^9,2$) and IrOA($25,5^6,2$). These arrangements generate 2-uniform states of 
homogeneous systems: 4 qutrits, 9 qubits and 6 ququints, respectively.
Two uniform states characterizing genuine heterogeneous systems are 
associated to difference schemes $D(N,N,d)$ 
such that $dN$ is not a power of a prime.
 
The simplest heterogeneous case arises from $D(6,6,3)$ which leads to IrMOA($18,3^63^12^1,2$)$\equiv$IrMOA($18,3^72^1,2$) 
and generates the 2-uniform state of the system $3^7\times2$:
\begin{eqnarray}\label{phi2137}
 &&|\phi_{3^7 2^1}\rangle=\nonumber\\
&&|00000000\rangle+|00112211\rangle+|01022120\rangle+\nonumber\\
&&|01201201\rangle+|02121010\rangle+|02210121\rangle+\nonumber\\
&&|11111100\rangle+|11220011\rangle+|12100220\rangle+\nonumber\\
&&|12012001\rangle+|10202110\rangle+|10021221\rangle+\nonumber\\
&&|22222200\rangle+|22001111\rangle+|20211020\rangle+\nonumber\\
&&|20120101\rangle+|21010210\rangle+|21102021\rangle.
\end{eqnarray}
In the same way, the difference scheme $D(10,10,5)$ implies the array IrMOA($50,5^{10}5^12^1,2$)$\equiv$IrMOA($50,5^{11}2^1,2$),
which  produces the state of the system $5^{11}\times2$:
\begin{widetext}
\begin{eqnarray}\label{phi21511}
|\phi_{5^{11} 2^1}\rangle&=&|000000000000\rangle+|001312243411\rangle+|012344012320\rangle+|013102434231\rangle+|020443231140\rangle+\nonumber\\
&&|024131302401\rangle+|031421420310\rangle+|032230144121\rangle+|043214321030\rangle+|044023113241\rangle+\nonumber\\
&&|111111111100\rangle+|112423304011\rangle+|123400123420\rangle+|124213040331\rangle+|131004342240\rangle+\nonumber\\
&&|130242413001\rangle+|142032031410\rangle+|143341200221\rangle+|104320432130\rangle+|100134224341\rangle+\nonumber\\
&&|222222222200\rangle+|223034410111\rangle+|234011234020\rangle+|230324101431\rangle+|242110403340\rangle+\nonumber\\
&&|241303024101\rangle+|203143142010\rangle+|204402311321\rangle+|210431043230\rangle+|211240330441\rangle+\nonumber\\
&&|333333333300\rangle+|334140021211\rangle+|340122340120\rangle+|341430212031\rangle+|303221014440\rangle+\nonumber\\
&&|302414130201\rangle+|314204203110\rangle+|310013422421\rangle+|321042104330\rangle+|322301441041\rangle+\nonumber\\
&&|444444444400\rangle+|440201132311\rangle+|401233401220\rangle+|402041323131\rangle+|414332120040\rangle+\nonumber\\
&&|413020241301\rangle+|420310314210\rangle+|421124033021\rangle+|432103210430\rangle+|433412002141\rangle.\nonumber\\
\end{eqnarray}
From $D(12,12,3)$ we generate IrMOA($36,3^{12} 3^1 2^1,2$)$\equiv$IrMOA($36,3^{13} 2^1,2$) and thus a state of the system $3^{13}\times2$,
\begin{eqnarray}\label{phi31322}
&&|\phi_{3^{13}2^1}\rangle=\nonumber\\
&&|00000000000000\rangle+|00001122112211\rangle+|00110011222220\rangle+|00112222001101\rangle+|01020212210110\rangle+\nonumber\\
&&|01022021121021\rangle+|01201201201200\rangle+|01202110022111\rangle+|02121210102020\rangle+|02122101010201\rangle+\nonumber\\
&&|02210120211010\rangle+|02211002120121\rangle+|11111111111100\rangle+|11112200220011\rangle+|11221122000020\rangle+\nonumber\\
&&|11220000112201\rangle+|12101020021210\rangle+|12100102202121\rangle+|12012012012000\rangle+|12010221100211\rangle+\nonumber\\
&&|10202021210120\rangle+|10200212121001\rangle+|10021201022110\rangle+|10022110201221\rangle+|22222222222200\rangle+\nonumber\\
&&|22220011001111\rangle+|22002200111120\rangle+|22001111220001\rangle+|20212101102010\rangle+|20211210010221\rangle+\nonumber\\
&&|20120120120100\rangle+|20121002211011\rangle+|21010102021220\rangle+|21011020202101\rangle+|21102012100210\rangle+\nonumber\\
&&|21100221012021\rangle.
\end{eqnarray}
\end{widetext}
There exist a useful compact way to write every
state arising from Construction \ref{C1} and \ref{C2}.
Introducing the shift operator,
\begin{equation}\label{GHZ4}
X|j\rangle=|[j+1]_d\rangle,
\end{equation}
the GHZ state of $N$ subsystems with $d$ levels each,
\begin{equation}
\label{GHZ2}
|GHZ_0\rangle=  \sum_{m=0}^{d-1} |m\rangle^{\otimes N},
\end{equation}
and its generalization,
\begin{equation}\label{GHZ2b}
|GHZ_j\rangle=\bigotimes_{k=0}^{N-1}X^{D(s,N,d)_{jk}}|GHZ_0\rangle|[k]_{p_1}\dots [k]_{p_{m^{\prime}}}\rangle,
\end{equation}
we can write 
\begin{equation}
\label{GHZ}
|\phi_{d^Np^1_1\dots p^1_m}\rangle=\sum_{j=0}^{s-1}|GHZ_j\rangle|[j]_{p_{m^{\prime}+1}}\dots [j]_{p_m}\rangle.
\end{equation}
where addition in Eq.(\ref{GHZ4}) is understood modulo $d$, $[k]_{p_i}=k\,(\mathrm{mod}\,p_i)$, $\{p_1,\dots,p_{m^{\prime}}\}$ are the distinct prime factors of $N$ and $\{p_{m^{\prime}+1},\dots,p_m\}$ is the subset of the distinct prime factors of $s$ which are not in common with the prime factors of $N$.
 
The superpositions of GHZ--like states (\ref{GHZ}) should not be confused with a particular class of mixed quantum states called $X$ \emph{states} \cite{YE07}. The $X$ states are convex combinations of GHZ--like rank-one projectors $|GHZ\rangle\langle GHZ|$,
 whereas here we consider superpositions of GHZ--like vectors $|GHZ_j\rangle$. 

\section{Endurance of k-uniformity}
In Section \ref{S_QS} we generated $k$-uniform states for heterogeneous systems 
by considering IrMOA. Here, we show that some columns of IrMOA can be
 removed without loosing the irredundancy of the orthogonal array in question.
 As consequence, from the $k$-uniform state $\{|\phi_{d^Np^1_1\dots p^1_m}\rangle\}$ we are able to generate the family of $k$-uniform states $\{|\phi_{d^xp^1_1\dots p^1_m}\rangle\}$, 
where 
\begin{equation}
N-\mu_k\leq x\leq N,
\end{equation}
and $m$ is the number of distinct primes factors of $sN$.
Here, the number $\mu_k$ is called \emph{endurance of k-uniformity} and it represents the maximal number of columns of an IrMOA that can be removed so that the remaining MOA preserves both irredundancy and strength $k$. It is important to stress that the $N-x$ columns to be removed have to contain exactly $d$ different symbols.  Otherwise, the endurance of $k$-uniformity would be not univocally defined. Observe that the action of removing a column in an IrMOA is not related to a local measurement performed by a given party. Hence, endurance of $k$-uniformity is not related to the \emph{persistency of entanglement} \cite{BR01}, which represents the minimal number of parties that should make a local measurement in order to assure that the remaining state is a fully separable. 

Indeed, by removing $N-2$ columns of the IrOA($2,2^N,1\mbox{)}=\{0\dots0;1\dots1\}$ we get the IrOA(2,$2^2,1\mbox{)}=\{00;11\}$. This operation can be associated to a mapping of the GHZ state of $N$ qubits into the Bell state. Thus, the IrOA(2,$2^N$,1) has endurance of 1-uniformity $\mu_1=N-2$, as 1-uniformity remains after removing $N-2$ columns of the IrOA. On the other hand, any local measurement on a system prepared in a GHZ state of $N$ parties lead us to a fully separable state, so the GHZ state has persistency of entanglement $P_E=1$ for any number of parties $N$. 

Table \ref{Table1} shows the endurance of $k$-uniformity for every state explicitly constructed in Section \ref{S_QS}. Let us highlight a 1-uniform the state appearing in this table. After removing the second and third column of the state $|\phi_{3^42^1}\rangle$ defined in Eq.(\ref{phi3421}) we generate the 1-uniform state
\begin{equation}\label{phi332}
|\phi_{3^22^1}\rangle=|000\rangle+|011\rangle+|120\rangle+|101\rangle+|210\rangle+|221\rangle.
\end{equation}
In Appendix \ref{AppendixA} we demonstrate that the simplest 1-uniform state for heterogeneous systems corresponds to the system $3\times 2\times 2$ and it does not arise from irredundant orthogonal arrays. This state is given by
\begin{eqnarray}\label{phi322}
|\phi_{3^12^2}\rangle&=&|0\rangle\bigl(|00\rangle+|11\rangle\bigr)+|1\rangle\bigl(|01\rangle+|10\rangle\bigr)+\nonumber\\
&&|2\rangle\bigl(|00\rangle-|11\rangle\bigr).
\end{eqnarray}
Note that this state is written in a linear decomposition of three orthogonal Bell states. Also, it is symmetric under the interchange of qubits. 

The simplest case for 2-uniform states of heterogeneous systems arises by removing two columns in the IrMOA associated to the state (\ref{phi2137}). Therefore, the following 2-uniform state for systems $3^5\times 2$ arises:
\begin{eqnarray}\label{phi2135}
|\phi_{3^5 2^1}\rangle&=&|000000\rangle+|001121\rangle+|010220\rangle+\nonumber\\
&&|012011\rangle+|021210\rangle+|022101\rangle+\nonumber\\
&&|111110\rangle+|112201\rangle+|121000\rangle+\nonumber\\
&&|120121\rangle+|102020\rangle+|100211\rangle+\nonumber\\
&&|222220\rangle+|220011\rangle+|202110\rangle+\nonumber\\
&&|201201\rangle+|210100\rangle+|211021\rangle.
\end{eqnarray}

In the on-line catalog of Kuhfeld \cite{catalog_Kuhfeld} there are several difference schemes $D(s,N,d)$ for $s>2$ which allow us to generate further 1-uniform and 2-uniform states 
of heterogeneous systems in the same way as Constructions \ref{C1} and \ref{C2}. 

Let us mention that 2-uniform states for $N$ qudits having a prime number of levels $d$ can be obtained with use of Construction \ref{C2} followed by removing the columns associated to the heterogeneous system. 

\begin{table}
\begin{tabular}{c|c|c|c||c|c}
$|\phi_{d_1^{n_1}d_2^{n_2}}\rangle$ &  \hspace{0.15cm}$k$\hspace{0.15cm} & \hspace{0.1cm}$\mu_2$\hspace{0.1cm} & \hspace{0.1cm}$\mu_1$\hspace{0.1cm} & $|\phi_{d_1^{n_1-\mu_2}d_2^{n_2}}\rangle$ & $|\phi_{d_1^{n_1-\mu1}d_2^{n_2}}\rangle$ \\ \hline\hline
$|\phi_{2^3}\rangle$ & 1 & - & 0 & - &  $|\phi_{2^3}\rangle$\\
$|\phi_{3^42^1}\rangle$ &1 & - & 2 & - & $|\phi_{3^22^1}\rangle$ \\
$|\phi_{5^62^1}\rangle$ & 1 & - & 4 & - & $|\phi_{5^22^1}\rangle$\\
$|\phi_{3^72^1}\rangle$ & 2 & 2 & 3 & $\mathbf{|\phi_{3^52^1}\rangle}$ & $|\phi_{3^42^1}\rangle$\\
$|\phi_{5^{11}2^1}\rangle$ & 2 & 6 & 7 & $|\phi_{5^{5}2^1}\rangle$ & $|\phi_{5^{4}2^1}\rangle$\\
$|\phi_{3^{13}2^1}\rangle$ & 2 & 6 & 8  & $|\phi_{3^{7}2^1}\rangle$ & $|\phi_{3^{5}2^1}\rangle$ \\
\end{tabular}
\caption{Endurances of uniformity $\mu_2$ and $\mu_1$ for every state $|\phi_{d_1^{n_1}d_2^{n_2}}\rangle$ explicitly presented in Section \ref{S_QS}. The first three rows correspond to 1-uniform states and the last three rows to 2-uniform states. The resulting 2-uniform and 1-uniform states obtained after removing $\mu_2$ and $\mu_1$ columns, respectively, are shown in the 4-\emph{th} and 5-\emph{th} column. The 2-uniform state for heterogeneous system belonging to the smallest possible Hilbert space that we found in this work is $|\phi_{3^52^1}\rangle$ (bold).}
\label{Table1}
\end{table}

\section{Quantum steering for heterogeneous systems}\label{Applic}
\begin{figure}
\begin{center}
\scalebox{0.33}{\includegraphics{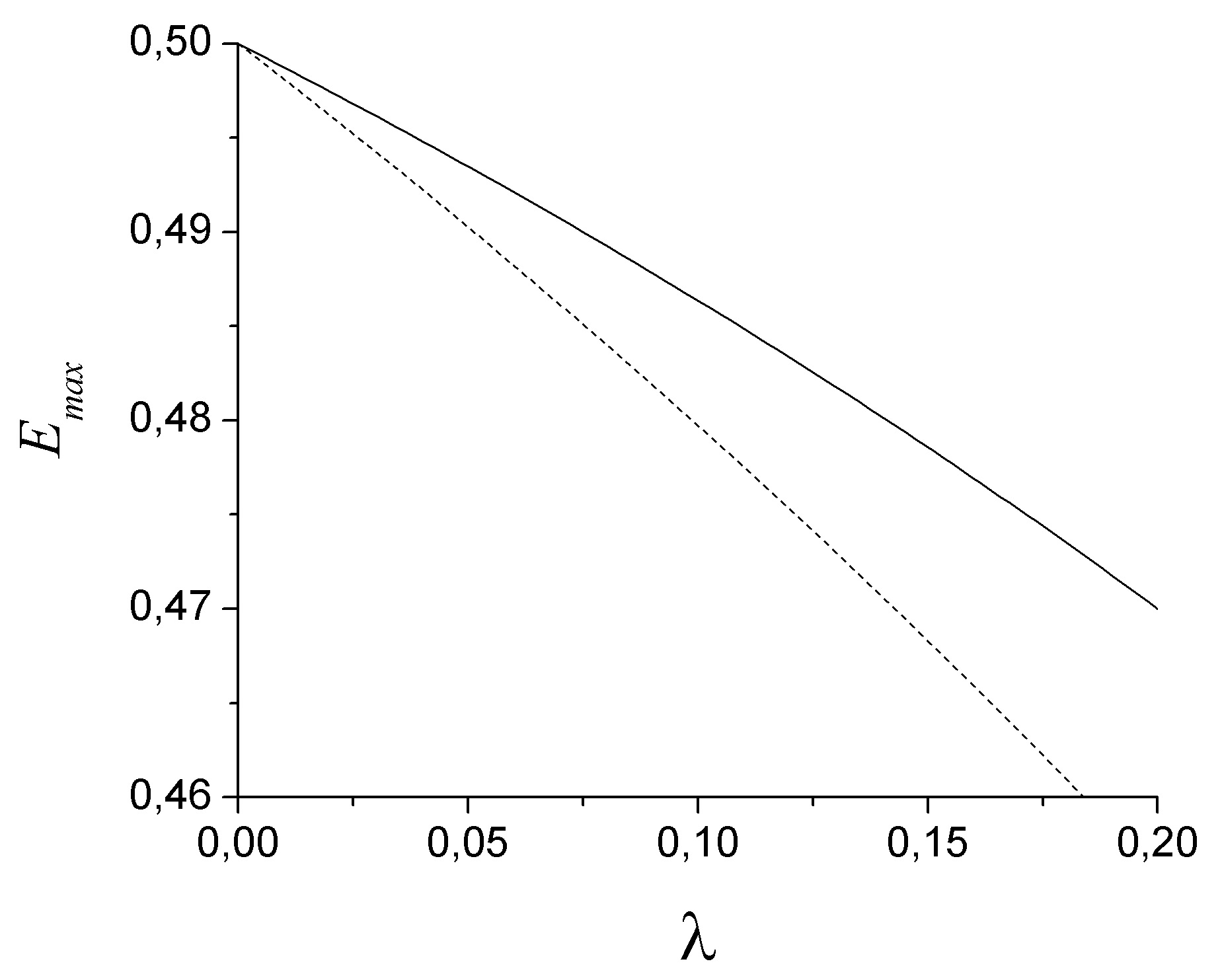}}
\end{center}
\caption{Maximal possible entanglement ($E_{max}$) for the bipartite system of Bob-Charlie when Alice steers the state under the presence of white noise characterized by the parameter $\lambda$ (see Eqs. (\ref{noiseGHZ}) and (\ref{noise322})). The solid lines corresponds to the heterogeneous system $3\times 2\times 2$ and the dashed line to the homogeneous system $2\times 2\times 2$. The qutrit allows to Alice to steer the state of Bob-Charlie in a more robust way, exhibiting a practical application of heterogeneous quantum states.}
\label{Fig1}
\end{figure}

In Section \ref{S_QS} we have found quantum states for heterogeneous systems exhibiting high multipartite entanglement. In this section, we show that the consideration of heterogeneous systems helps us to implement quantum steering in a more efficient way. For the system $3\times 2\times 2$ we have found the 1-uniform state
$|\phi_{322}\rangle$, given in Eq.(\ref{phi322}). This state has interesting properties in common with the GHZ state of three qubits. Indeed, for a system $3\times 2\times 2$ prepared in the state $|\phi_{322}\rangle$ Alice is able to steer the state of Bob-Charlie from a separable to a maximally entangled one. Specifically, if Alice collapses her subsystem into the state $|\psi_A\rangle=|0\rangle+|2\rangle$ then Bob and Charlie are in the fully separable state $|00\rangle$. On the other hand, if Alice collapses her state to $|\psi'_A\rangle=|0\rangle$ then the remaining state of Bob-Charlie is the Bell state $|00\rangle+|11\rangle$. Alice is also able to steer the state of Bob-Charlie in a similar way when she has a \emph{qubit} insted of a \emph{qutrit} if the system is prepared in the state $|GHZ\rangle$ \cite{HR13}. Thus, concerning steering there is no apparent advantage of the third extra level of Alice provided by the qutrit. However, there is a clear advantage for realistic implementations, i.e., under the presence of noise. In Fig.\ref{Fig1} we compare the robustness of entanglement of the steered state of Bob-Charlie under the presence of white noise for the cases: \emph{(i)} the system is $2\times 2\times 2$ and the state is $|GHZ\rangle$ \emph{(ii)} the system is $3\times 2\times 2$ and the state is $|\phi_{322}\rangle$. To this end, we consider the noisy tripartite states
\begin{equation}\label{noiseGHZ}
\rho_{\lambda}^{GHZ}=(1-\lambda)|GHZ\rangle\langle GHZ|+\lambda\,\mathbb{I}_{8},
\end{equation}
and
\begin{equation}\label{noise322}
\rho_{\lambda}^{322}=(1-\lambda)|\phi_{322}\rangle\langle \phi_{322}|+\lambda\,\mathbb{I}_{12},
\end{equation}
respectively. As we can see from Fig.\ref{Fig1}, the fact that Alice posses a \emph{qutrit} instead of a \emph{qubit} leads to more robustness of entanglement under the presence of noise for the subsystem of Bob-Charlie. As measure of entanglement, we considered here the \emph{negativity} \cite{ZHSL98}, i.e., the absolute value of the sum of the negative eigenvalues of the matrix $\rho^{T_2}_{AB}$, where $T_2$ denotes the partial transposition. A realistic value of the fidelity for the experimentally generated GHZ state of three qubits is $F=\langle GHZ|\rho_{\lambda}^{GHZ}|GHZ\rangle\approx0.89$ (trapped ions \cite{R04}), which means that $\lambda\approx0.13$, under the assumption of a white noise model. Therefore, the advantage of the qutrit over the qubit on Alice side implies an additional 2\% of entanglement (see Fig. \ref{Fig1}). Despite this simple illustration does not have an important practical advantage the robustness of entanglement increases for a higher number of particles and levels, where entanglement is much more sensitive to the presence of noise.

Finally, let us recall that the 1-uniform and 2-uniform states for systems $d^N\times p_1\times\dots\times p_m$ allow the general form (see Eq.(\ref{GHZ}))
\begin{equation*}
|\phi_{d^Np^1_1\dots p^1_m}\rangle=\sum_{j=0}^{s-1}|GHZ_j\rangle|[j]_{p_{m^{\prime}+1}}\dots [j]_{p_m}\rangle.
\end{equation*}
This implies that any single party associated to a subsystem having $p_k$ levels can steer the state of the homogeneous part of $N$ qudits to $p_k$ mutually orthogonal GHZ states. For example, the state of the system $3\times 3\times 2$ defined in Eq.(\ref{phi332}) is equivalent to the state
\begin{eqnarray}
&&|\phi_{3^2 2^1}\rangle=\\
&&\bigl(|00\rangle+|11\rangle+|22\rangle\bigr)|0\rangle+\bigl(|01\rangle+|12\rangle+|20\rangle\bigr)|1\rangle.\nonumber
\end{eqnarray}
Here, Charlie can steer the state of Alice-Bob to the orthogonal Bell-like states:
\begin{equation}
|\psi_{3^2}\rangle_{AB}=|00\rangle+|11\rangle+|22\rangle,
\end{equation}
and
\begin{equation}
|\psi^{\prime}_{3^2}\rangle_{AB}=|01\rangle+|12\rangle+|20\rangle,
\end{equation}
if he collapses his subsystem into the states $|\psi_C\rangle=|0\rangle$ and $|\psi'_C\rangle=|1\rangle$, respectively.

\section{Concluding remarks}
Multipartite quantum states were usually studied for
homogeneous systems consisting of subsystems
with the same number of levels.
As several experimentally studied physical systems 
consists of subsystems of different numbers of levels
in this work we analyzed quantum entanglement in 
heterogeneous systems.

Highly entangled quantum states of homogeneous
systems consisting of $N$ subsystems with $d$ levels each
can be related with irredundant orthogonal arrays \cite{GZ14,FJXY15} 
containing symbols from a $d$-letter alphabet.
Making use of mixed orthogonal arrays \cite{Rao,HPS92}
we generalized this relation for quantum states
of heterogeneous systems.

In particular, we presented explicit 
constructions of one and two--uniform states  
for several heterogeneous quantum systems.
Simple cases include the one uniform state
(\ref{phi332})
of a system consisting of two qutrits and one qubit and (\ref{phi322}) for one qutrit and two qubits. Further cases are the one uniform state
(\ref{phi3421}) of four qutrits and one qubit,
and state  (\ref{phi5521})
for six subsystems with five levels each and one qubit.
Furthermore, the state (\ref{phi2137}) of a system containing seven qutrits and one qubit is two uniform. From here, the state of 5 qutrits and 1 qubit (\ref{phi2135}) arises, which represent the simplest 2-uniform state presented in this work (see Table \ref{Table1}).

It is tempting to believe that some states discussed in this work can be useful for experimental purposes, as they exhibit more robustness under noise for multipartite quantum steering. Furthermore, the multipartite heterogeneous states minimize the number of levels required by a single party to steer the state of its complementary part to maximally entangled states (see end of Section \ref{Applic}).

Finally, let us mention that $k$--uniform states for heterogeneous systems can be useful to construct quantum error correction codes. Several questions related to maximally entangled states for heterogeneous systems remain open. For example, it is not known for which heterogeneous quantum systems $k$--uniform states exists. In particular, it would be interesting to find examples of $3$--uniform states for such systems.

\medskip

{\bf Acknowledgements}.
It is a pleasure to thank P. Horodecki, M. Grassl and A. Sudbery for useful remarks. This work has been supported by the Polish National Science Center under the project number DEC-2011/02/A/ST1/00119, by the European Union FP7 project PhoQuS@UW (Grant Agreement No. 316244) and by the John Templeton Foundation under the project No. 56033.

\appendix

\section{k-uniform states from redundant MOA}\label{AppendixA}
For some systems there exist $k$-uniform states 
which are not related to mixed irredundant orthogonal arrays.
Such examples were discussed for standard OA, where some minus signs had to be introduced in order to generate 
certain  entangled states for qubits -- see Appendix C in Ref. \cite{GZ14}.
Here, we construct a 1-uniform state for heterogeneous systems following the same procedure.

Let us start by considering the state of a system $3\times2\times2$:
\begin{eqnarray}
|\phi_{3^1 2^2}\rangle&=&|000\rangle+e^{i\alpha_1}|011\rangle+e^{i\alpha_2}|101\rangle+\nonumber\\
&&e^{i\alpha_3}|110\rangle+e^{i\alpha_4}|200\rangle+e^{i\alpha_5}|211\rangle.\nonumber
\end{eqnarray}
Here, the corresponding array is MOA but not irredundant. Indeed, an IrMOA($r,3^12^2,1$) do not exist for any $r$. For this state, the reductions to the second and third party are maximally mixed for any value of the parameters $\{\alpha_j\}$. That is, 
\begin{eqnarray}
\rho_B&=&\mathrm{Tr}_{AC}(|\phi_{3^2 2^1}\rangle\langle\phi_{3^2 2^1}|)=\frac{1}{2}\mathbb{I}_2,\nonumber\\
\rho_C&=&\mathrm{Tr}_{AB}(|\phi_{3^2 2^1}\rangle\langle\phi_{3^2 2^1}|)=\frac{1}{2}\mathbb{I}_2.\nonumber
\end{eqnarray}
This is so because after removing the second or third column the remaining MOA of two columns is irredundant. However, after removing the first column there are repetition of symbols. The only restriction to have $\rho_A$ maximally mixed is given by
\begin{equation}\label{state233}
e^{i (\alpha_1 - \alpha_5)} + e^{-i\alpha_4}=0,
\end{equation}
where $\alpha_1=\dots=\alpha_4=0$ and $\alpha_5=\pi$ is a solution. The fact that there is only one equation for the phases $\alpha_j$ implies that there is only one redundancy for the MOA. Therefore, the following state of system $3\times2\times2$ is 1-uniform:
\begin{equation}\label{phi3221_2}
|\phi_{3^1 2^2}\rangle=|000\rangle+|011\rangle+|101\rangle+|110\rangle+|200\rangle-|211\rangle.\\
\end{equation}
To our best knowledge, the maximally entangled state (\ref{phi3221_2}) did not appear in the literature so far.  

\section{IrOA from Galois fields}\label{AppendixB}
The catalog of difference schemes \cite{catalog_Kuhfeld} contains the following case:
\begin{equation}
D(4,4,4)=\begin{array}{cccc}
0&0&0&0\\
0&1&2&3\\
0&2&3&1\\
0&3&1&2
\end{array}.
\end{equation}
In order to construct the IrOA(16,4,4,2) from D(4,4,4) we have to consider the Galois field $GF(4)$. The elements of this field are given by $0, 1, X$ and $1+X$, which are represented in the scheme as $0,1,2,3$, respectively. The $16\times4$ arrangement is given by
\begin{eqnarray}
&&D(4,4,4)\nonumber\\
&&D(4,4,4)+1\nonumber\\
&&D(4,4,4)+X\nonumber\\
&&D(4,4,4)+1+X,
\end{eqnarray}

\begin{table}[h]
\begin{tabular}{c|cccc}
+&0&1&X&1+X\\ \hline
0&0&1&X&1+X\\
1&1&0&1+X&X\\
X&X&1+X&0&1\\
1+X&1+X&X&1&0
\end{tabular}
\caption{Addition table for elements of the Galois Field $GF(4)$.}
\label{Table2}
\end{table}
\noindent where addition of the elements of $GF(4)$ is given according to Table \ref{Table2}. Therefore, from Construction \ref{C2} we have the orthogonal array IrOA(16,4,4,2)$^t$ given by
\begin{equation}
\begin{array}{cccccccccccccccc}
0&0&0&0&1&1&1&1&2&2&2&2&3&3&3&3\\
0&1&2&3&1&0&3&2&2&3&0&1&3&2&1&0\\
0&2&3&1&1&3&2&0&2&0&1&3&3&1&0&2\\
0&3&1&2&1&2&0&3&2&1&3&0&3&0&2&1
\end{array},
\end{equation}\vspace{0.5cm}

\noindent where $t$ denotes transposition. The corresponding 2-uniform state is given by
\begin{eqnarray}
|\phi_{4^4}\rangle\!\!&=\!\!&|0000\rangle+|0123\rangle+|0231\rangle+|0312\rangle+\nonumber\\
&&|1111\rangle+|1032\rangle+|1320\rangle+|1203\rangle+\nonumber\\
&&|2222\rangle+|2301\rangle+|2013\rangle+|2130\rangle+\nonumber\\
&&|3333\rangle+|3210\rangle+|3102\rangle+|3021\rangle.
\end{eqnarray}

\end{document}